\documentclass[USenglish,twocolumn]{article}

\usepackage[utf8]{inputenc}
\usepackage[big]{dgruyter_author}

\begin{document}

  \articletype{Research Article{\hfill}Open Access}

  \author*[1]{Olivier EMILE}

\author[2]{Janine EMILE}

  \affil[1]{Universit\'e de Rennes 1, 35042 Rennes cedex, France, E-mail: olivier.emile@univ-rennes1.fr}

  \affil[2]{IPR, UMR CNRS 6251, Universit\'{e}  Rennes 1, 35042 Rennes cedex, France, E-mail: janine.emile@univ-rennes1.fr}

  \title{\huge Towards an optofluidic pump? }

  \runningtitle{Towards an optofluidic pump?}


  \begin{abstract}
{Most of the vibrating mechanisms of optofluidic systems are based on local heating of membranes that induces liquid flow. We report here a new type of diaphragm pump in a liquid film based on the optical radiation pressure force. We modulate a low power laser that generates, at resonance, a symmetric vibration of a free standing soap film. The film lifetime strongly varies from 56 s at low power (2 mW) to 2 s at higher power (70 mW). Since the laser beam only acts mechanically on the interfaces, such a pump could be easily implemented on delicate micro-equipment on chips or in biological systems.}
\end{abstract}

  \journalname{ESJ physics}

  \startpage{1}

\maketitle
\section{Introduction}
At the photon level, the value of the linear momentum depends on the refractive index of the medium. At an interface, with two media with different indexes, for example at an air/water interface, the linear momentum is discontinuous. This leads to a radiation pressure force \cite{ashkin,casner2,loudon} that may then induce an interfacial deformation. However, in order to bend it, one has to compensate for the surface tension. Apart from experiments that use evanescent waves to act on the interface \cite{emile,emile2,emile3}, there are two main ways to compensate for it. On the one hand, one can apply high power focalized laser beams \cite{ashkin,ashkin2,astrath,klein} or, on the other hand, one can use two liquids with similar surface tensions \cite{casner,casner2}. Anyway, from a mechanical standpoint, it is easier to deform a liquid film under illumination than an interface. Besides the liquid film may be considered as a model case. It is therefore essential to determine the parameters of the parametric amplification excitation of the system that may lead to resonant vibrations. The aim of this letter is to investigate the lifetime of a soap film irradiated by a low power chopped laser and look for potential applications. 

\section{Experimental procedures}

\begin{figure}
\includegraphics[width=3.3In]{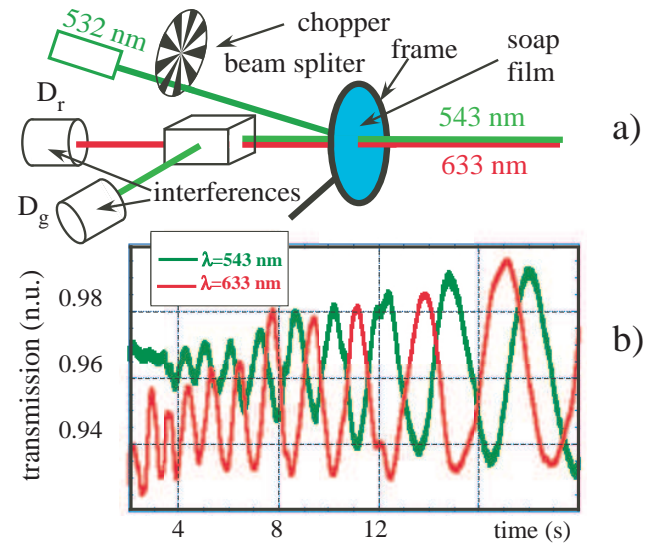}
  \caption{Experimental set up: a) a 532 nm modulated laser light deforms a vertical free standing thin soap film. The thickness is measured from the transmission interferences of two  low power lasers. BS: beam splitter, Dg and Dr: photodiodes. b) example of interference signals.}
  \label{fig1}
\end{figure}

The experimental set-up, already described in \cite{emile4}, is shown in Fig.~\ref{fig1}a. A chopped green laser (Cristal Laser, P~=~100~mW, that can be attenuated to P~=~2~mW, waist w~=~300~$\mu$m, $\lambda=532$~nm) illuminates a vertical free standing draining film supported by a glass frame. From the relative position of the intensity extrema of the interferences in transmission from two continuous (CW) He-Ne lasers (Melles Griot, P~=~1~mW, w~=~400~$\mu$m,  $\lambda=633$~nm and $\lambda=543$~nm), one measures the absolute film thickness \cite{emile5} with a precision better than 2~nm. The chopped laser and the two probing beams are superposed on the centre of the soap film. The vertical free draining film (laser off) has a lifetime of $27\pm1$ s. We have checked that when the solid state laser is not modulated nor applied, and when the probe lasers are off, the film lifetime is unchanged. The two He-Ne lasers have thus no visible effect on the film.

 \begin{figure}
\includegraphics[width=3.3In]{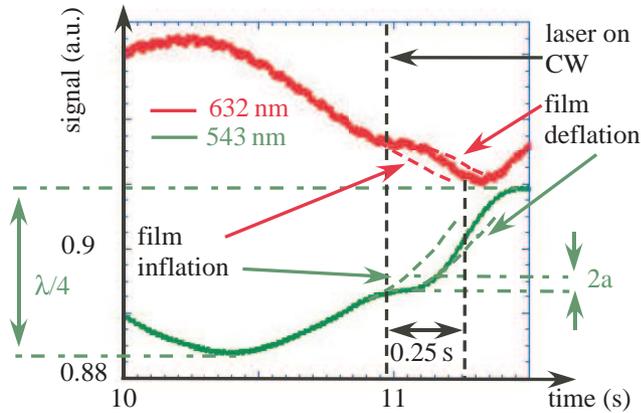}
  \caption{Dynamical response of the interferences of the probe lasers when the laser is switched on. The dotted lines correspond to the interferences without the laser. For the sake of clarity, the interference fringes have not been superimposed.}
  \label{fig2}
\end{figure}

The glass frame sustaining the film is a $d=3$~cm-diameter toroidal ring made from a 5~mm-diameter glass rod. The soap solution is made of SLES surfactant (Sodium Lauryl Ether Sulfate, Cognis), $0.1\%$ v/v diluted in pure water, below the critical micellar concentration (CMC). The temperature is controlled to T~$= 20.0\pm 0.2^{\circ}$C, as well as the air humidity to $50\%\pm 5\%$. The intensities of the two transmitted probe lasers are recorded on two photodiodes and registered on a computer (see Fig.~\ref{fig1}b) with an acquisition rate of 1 ms or 100 ms respectively. 

\section{Results}

\subsection{Dynamical response of the film.}

We have first registered the dynamical response of the film when suddenly illuminated by a CW laser (see Fig.~\ref{fig2}). One notes that the soap film inflates when the laser is switch on, as expected. This inflation $2a$ is the variation of the signal with the laser off (dotted line) and with the laser on (plain line). It can be evaluated by computing the transmission signal versus film thickness that leads to interferences. This establishes a correspondence between the detected signal and the film thickness  \cite{emile5}. This then enables the evaluation of the $2a$  variation that equals 14 nm in our case, both with the green and the red lights. It is small compared with the 1 $\mu$m film thickness. Thus each interface is bent (bump), with a maximum amplitude of 7 nm. It then relaxes in a single oscillation, meaning that the damping is strong. The period of this damped oscillation is estimated to 0.25 s (see Fig.~\ref{fig2}) which corresponds to a pseudo frequency of 4 Hz. We have made the same observations when the laser is suddenly switched off with a dip on both interfaces.  

\begin{figure}
\includegraphics[width=3.3In]{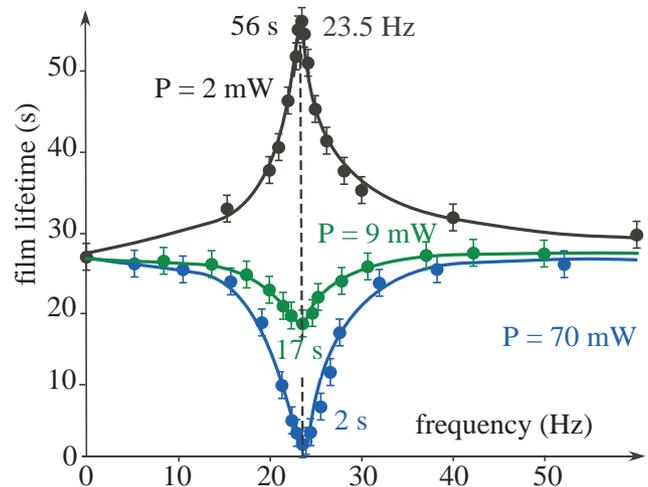}
\caption{Film lifetime versus modulation frequency for three different powers, showing a resonance at 23.5 Hz.}
 \label{fig3}
\end{figure}

\subsection{Resonance}

Since the system is strongly damped, one would expect a resonance frequency higher than the 4 Hz frequency, i.e. in the tenth of hertz range. We have then modulated the laser (2 mW) with different frequencies. The film lifetime versus the modulation frequency is displayed in Fig.~\ref{fig3}. Every measurement is done at least three times, with a very good reproducibility. At low power (2 mW), the curve exhibits a sharp resonance at a frequency of $23.5\pm0.1$ Hz. One also notes that when the light is not modulated (CW laser), the lifetime is unchanged. At high frequency, the lifetime is also unchanged (27 s). 

As the power is further increased, the resonance is continuously turned into an antiresonance. At a power of 70 mW, the film lifetime is only two seconds. When illuminated with a chopped light at 23.5 Hz, the film nearly immediately breaks. This is not due to heating and thermal effects since,  with CW light, even at high power, the film lifetime is unchanged. It is only because the amplitude of the oscillations increases with the power that the film becomes brittle. Actually, we have not measured the amplitude of the oscillation. The dynamical response of the film under a single 2 mW excitation is 7 nm. At resonance, the amplitude of the oscillation must be higher. Assuming a linear dependance of the amplitude with the applied power, the amplitude of the oscillation would be of the order of hundreds of nm at 70 mW, i.e. close to the film thickness. This explains why the film breaks instantaneously. 

It has to be noted that the resonance or antiresonance frequency doesn't depend on the optical power in the explored power range. It hardly varies with the concentration, or the temperature, or the frame geometry. We have also checked that this resonance phenomenon doesn't depend on the wavelength of the laser, leading always to a resonance frequency of 23.5 Hz. This is also true with white light. 

\section{Discussion}

When a force is applied on the air/liquid interface the surface tension acts as a spring. Then the oscillation frequency $\nu$ of the film should be of the order of
\begin{equation}
\nu=1/2\pi(\alpha/m)^{1/2}
 \label{eq1}
\end{equation} 
$\alpha$ being the surface tension coefficient, $m$ being homogeneous to a mass. Here, the film mass is about $7\times10^{-7}$ kg. With $\alpha=0.03$ Nm$^{-1}$ \cite{SLES}, one finds $\nu=30$ Hz. This is the correct order of magnitude for the resonance we found. However, this is just an estimation of it. A two times higher mass for example, according to equation~\ref{eq1}, would lead to a 21 Hz resonance frequency, closer to the value we found. The resonance frequency probably depends on the film surface and on its chemical properties. For example, the interfacial visco-elasticity of the liquid film can be studied by its mechanical response, with exchanges between surface and bulk from the film and also from the meniscus between the film and the frame. Moreover, it is difficult to increase significantly the surface tension of a liquid film, however vibrating a fiberglass with a higher surface tension could to be an alternative.

The film indeed vibrates on a symmetric mode (also called squeezing mode), because the laser bends each interface in a symmetrical way. The film inflates when the laser is on and deflates when off, like in a breathing mode. It is different from the asymmetric modes (also called bending mode) usually seen in soap films under acoustic excitation where both interfaces are bended in an asymmetric way \cite{couder,elias,ben}. Here, we can't excite bending modes, we only excite breathing modes which may have more practical applications. The liquid within the film is pumped in and out when the laser is modulated. Even at very low power (2 mW), with natural light, the film is vibrating like a membrane in a diaphragm pump. The liquid flow is in the nano-liter range. It could be precisely adjusted by changing the laser power. 

Connecting the frame with two reservoirs, using microfluidic valves as the ones that have been demonstrated recently \cite{valve1}, the film would behave as an optofluidic pump. Since it vibrates at resonance, the optical power could be very low which may be a crucial issue in biological systems where optical power is detrimental to cells in general \cite{scackmann}. Such a system would be cheap, and even sun light can be used. Moreover, since it might be difficult to make a thin film from any liquid, this film can be enclosed in a soft transparent membrane as soon as this membrane is soft enough to vibrate under light excitation. The resonance frequency may then be different from the low resonance frequency we found here. 

\section{Conclusion}

We have shown that even a low power laser can dramatically change the dynamical behavior of a liquid film. At low power, a modulated laser beam enhanced the lifetime of a thin soap by a factor of two, whereas at higher power, the film breaks nearly instantly. The film vibrates in a symmetric mode and behaves as a vibrating membrane in a diaphragm pump. 

Such a mechanical action on a liquid interface at resonance without any invasive operation could also find issues in biology such as in the-in-or-out pumping of liquids in biological cells. It may also be of timely interest in dermatology to remove angioma, in opthalmology to distroy floaters, or in cataract operation. 

\paragraph*{Acknowledgments}
The authors thank J.~R.~Th\'ebaut for technical help. This work has been performed within the EU COST action MP 1205 Advances in Optofluidics: integration of optical control and photonics with microfluidics.

\end{document}